# Witnessing the formation and relaxation of massive quasi-particles in a strongly correlated electron system

Fabio Novelli[1,2], Giulio De Filippis[3]*, Vittorio Cataudella[3], Martina Esposito[1], Ignacio Vergara Kausel[4], Federico Cilento[2], Enrico Sindici[1], Adriano Amaricci[5], Claudio Giannetti[7,8], Dharmalingam Prabhakaran[9], Simon Wall[10], Andrea Perucchi[2], Stefano Dal Conte[6], Giulio Cerullo[6], Massimo Capone[5], Andrey Mishchenko[11,12], Markus Grüninger[4], Naoto Nagaosa[11,13], Fulvio Parmigiani[1,2], and Daniele Fausti*[1,2]

(1) Dipartimento di Fisica, Università degli Studi di Trieste, 34127 Trieste, Italy

(2) Sincrotrone Trieste S.C.p.A., 34149 Basovizza, Italy

(3) SPIN-CNR and Dipartimento di Scienze Fisiche, Università di Napoli Federico II, I-80126, Napoli, Italy

(4) Department of Physics, University of Cologne, 50923 Köln, Germany

(5) CNR-IOM Democritos National Simulation Center and Scuola Internazionale Superiore di Studi Avanzati (SISSA), Via Bonomea 265, 34136 Trieste, Italy

(6) IFN-CNR, Dipartimento di Fisica, Politecnico di Milano, Piazza L. da Vinci 32, 20133 Milano, Italy

(7) I-LAMP (Interdisciplinary Laboratories for Advanced Materials Physics), Università Cattolica del Sacro Cuore, Brescia I-25121, Italy

(8) Department of Physics, Università Cattolica del Sacro Cuore, Brescia I-25121, Italy

(9) Department of Physics, University of Oxford, Oxford, OX1 3PU, UK

(10) ICFO-Institut de Ciencies Fotoniques, Av. Carl Friedrich Gauss, 3, 08860 Castelldefels, Barcelona, Spain

(11) RIKEN Center for Emergent Matter Science (CEMS), Wako 351-0198, Japan

(12) RRC "Kurchatov Institute", 123182, Moscow, Russia

(13) Department of Applied Physics, The University of Tokyo, 7-3-1 Hongo, Bunkyo-ku, Tokyo 113, Japan

* Correspondence to: daniele.fausti@elettra.trieste.it; giuliode@na.infn.it

**The non-equilibrium semiconductors physics is based on the paradigm that different degrees of freedom interact on different timescales. In this context the photo-excitation is often treated as an impulsive injection of electronic energy that is transferred to other degrees of freedom only at later times. Here, by studying the ultrafast particles dynamics in a archetypal strongly correlated charge-transfer insulator ($La_2CuO_4$), we show that the interaction between electrons and bosons manifest itself directly in the photo-excitation processes of a correlated material. With the aid of a general theoretical framework (Hubbard Holstein Hamiltonian), we reveal that sub-gap excitation pilots the formation of itinerant quasi-particles which are suddenly dressed (<100 fs) by an ultrafast reaction of the bosonic field.**

The exotic electronic ground states of transition metal oxides (TMOs), such as high temperature superconductivity and giant magnetoresistance, arise from both the strong interactions among electrons and the complex interplay between the electronic and other bosonic degrees of freedom such as vibrations and magnetic excitations[1-4]. A paradigmatic example of such complexity is represented by the cuprate family. Here, starting from parent compounds which are charge-transfer (CT) insulators[5-13], anomalous metallic and eventually high temperature superconducting phases appear upon electron or hole doping[14,15].

In CT insulators the most essential properties are determined by Coulomb repulsion which forces the electrons to localize, freezing charge fluctuations. In such a state the coupling between localized fermions and bosonic excitations, which is mediated by charge dynamics, are naively expected to be strongly inhibited[16,17].

While this scenario is seemingly met when we measure and study the properties of a stady-state Mott insulator, a different pattern can take place if the system is doped[18,19] or driven out of equilibrium[13,20]. In particular the naive picture would suggest that, if we try to excite the system by means of a pulse with photon energy smaller than the charge-transfer gap, we would observe little effects on the material as the localized carriers have no chance to change their state and, therefore, they have no means to couple to bosonic excitations. In our study a completely different scenario emerges, in which a coupled boson-charge mode is excited leading to an effective renormalization of the charge-transfer gap[21].

We show here that a model including both Coulomb interaction and coupling with a boson mode (Hubbard-Holstein Hamiltonian) can reproduce our experimental outcomes. Please note that in the following we will keep using the general term boson. This is due to the fact that, besides the Hubbard-Holstein Hamiltonian was developed in order to treat lattice vibrations, such model can be used to address in the same way other kinds of bosonic excitations like an electronic collective mode[22]. The effects of the electron-boson coupling are highlighted here by comparing the dynamical optical responses driven by resonant and off-resonant excitations of the CT absorption peak (CTP), i.e. with photon energy respectively larger and smaller than the energy needed to create free charge excitations (see Fig.1).

**Results**

From the measured $\Delta R(\omega,\tau)/R(\omega)=R(\omega,\tau)/R(\omega)-1$ with $R(\omega,\tau)$ the time-dependent pump-perturbed reflectivity and $R(\omega)$ the equilibrium reflectivity we calculate the pump-induced changes of the optical conductivity $\Delta\sigma_1(\omega,\tau)=\sigma_1(\omega,\tau)-\sigma_1(\omega)$ as a function of probe frequency ($\omega$) and pump-probe delay time ($\tau$). $\sigma_1(\omega,\tau)$ is the time dependent light-perturbed optical conductivity and $\sigma_1(\omega)$ is the stady-state optical conductivity. $\Delta\sigma_1(\omega,\tau)$ is obtained from $\Delta R(\omega,\tau)/R(\omega)$ with a novel data analysis methodology based on the Kramers-Kroenig relations. In short, by knowing the equilibrium reflectivity $R(\omega)$ and the non-equilibrium $\Delta R(\omega,\tau)/R(\omega)$ it is possible to calculate the pump-perturbed reflectivity $R(\omega,\tau)$ and, from it, all the transient optical properties (see Methods for details). The results are reported in Fig.2a and Fig.2b for pump pulses with photon energy at 3.1±0.04 eV and 0.95±0.04 eV, respectively. In the first hundred femtoseconds we observe striking differences between the photo-excitation with photons at energy $E_{pump}>E_{CTP}$ (Fig.2a) and $E_{pump}<E_{CTP}$ (Fig.2b). A smooth rise time (>100 fs) is detected only for pump energy above the CTP, while sub-CTP excitation leads to a pulse width limited rise time of the signal followed by an ultrafast decay. This difference is made evident in Fig.2c, where the optical response at the charge-transfer absorption peak (corresponding to about 2.3 eV) is shown as a function of the pump-probe delay. The results in Fig.2c are normalized to the energy absorbed by the sample for 3.1±0.04 eV (solid blue curve) and 0.95±0.04 eV pump pulses (solid red curve), see Methods for details.

In Fig.2c we show that the modification of the optical response at the charge-transfer edge to the two different stimula differs in the initial dynamics while it becomes similar for longer delays between the pump and the probe pulses ($\tau > \tau_2^{exp} = 200\,fs$). Such "slow" response is consistent with previous time-resolved measurements on $La_2CuO_4$[23,24], $Nd_2CuO_4$[23-25], and $YBa_2Cu_3O_Y$[25].

In order to address the pulse width limited rise time for sub-CTP excitation we performed time-resolved reflectivity measurements with single-colour probe pulses at 2.3±0.3 eV, near-infrared (visible) pump pulses at 1.4±0.2 eV (2.3±0.3 eV), and better temporal resolution (~15 fs). The transient optical reflectivity at the charge-transfer edge for near-infrared (NIR) pump pulses exhibits an ultrafast rise time that is again within the experimental resolution (inset of Fig.2c). This result indicates that the important physical parameter that governs the rise time is the pump photon energy, i.e. "fast" (<100 fs) or "slow" (>100 fs) optical responses are driven by pump pulses with energy respectively smaller or larger than the unperturbed CTP energy value.

We now take advantage of the broadband probe and describe the spectral dependence of the transient response. The energy dependent $\Delta\sigma_1(\omega,\tau)$ at $\tau$=0.05±0.05 ps ($\tau$=0.4±0.1 ps) is reported in Fig.3a (Fig.3c) for the two pump energies (blue for 3.1±0.04 eV and red for 0.95±0.04 eV pulses, respectively). For $E_{pump}>E_{CTP}$ at both pump-probe delays (blue curves in Fig.3a and Fig.3c) we observe a loss of spectral weight[26] (SW) centred at $E_{probe}$~2.3 eV and a gain of SW around $E_{probe}$~2.1 eV. This behaviour can be ascribed to a "thermal response", i.e. it is compatible with the shift towards lower energies of the CTP upon heating[23-29]. This is confirmed by comparing the measured $\Delta\sigma_1(\omega,\tau)$ at long pump-probe delays with the expected variation of the equilibrium optical conductivity for a temperature increase corresponding to the pump-induced heating, black line in Fig.3c (see Methods for details).

On the contrary $\Delta\sigma_1(\omega,\tau)$ for NIR pump, displayed in the red curves in Fig.3a and Fig.3c, reveals a significant spectral evolution with time. The main feature remains a loss of SW in the CTP region, but at different energies as a function of the pump-probe delay times. The $\Delta\sigma_1(\omega,\tau)$ peaks at $E_{probe}$~2.1 eV within $\tau_1^{exp} = 100\,fs$ and drifts towards the thermal response only on a longer times.

It is important to notice that for $\tau \leq \tau_1^{exp} = 100\,fs$ the optical conductivity displays only a negative variation around $E_{probe}$~2.3 eV, that corresponds to the energy edge of the unperturbed charge-transfer transition. Hence for $\tau \leq \tau_1^{exp}$ and $E_{pump}$=0.95±0.04 eV the main photo-induced effect consists of an overall quench of the CT transition. In order to rationalize these evidences we introduce in the next section the Hubbard-Holstein Hamiltonian and perform proper many-body calculations.

## Discussion

In a one-band Hubbard picture the optical gap for free-charge excitation is related to the energy cost for unbound double occupancies. In order to explain the differences between the responses to below and above CTP excitations one has to consider the two limit cases predicted by time-dependent perturbation theory: inelastic (non-adiabatic) and elastic (adiabatic) regimes. The non-adiabatic regime describes real transitions between electronic quantum states where the response is present for some time after the pump pulse is over. On the contrary in the adiabatic regime the response comes from the deformation of the wave functions by the electric field of the light and the system comes back to the ground state when the pump pulse is over. We argue that in our experiment, being the spectral-response almost independent on the pump-probe delay and lasting several picoseconds, the non-adiabatic contribution with real electronic transitions dominates above-CTP absorption (Fig.2a), while for $E_{pump}<E_{CTP}$ a more composite scenario emerges.

In order to describe the difference between the responses driven in $La_2CuO_4$ by sub-CTP, $E_{pump}=0.95\pm0.04$ eV, and above-CTP, $E_{pump}=3.1\pm0.04$ eV, pump pulses (Fig.2) we turned to the Hubbard-Holstein Hamiltonian (HHH) and introduce the effects of the electromagnetic fields through the classical vector potential $\mathbf{A}(\tau)$ (see Methods). The HHH represents a standard modeling for undoped cuprates where the insulating state arises from strong repulsion between electrons and the contribution of oxygen p orbitals are not treated explicitly but through the introduction of an effective repulsion U[30]. By taking into account both onsite Coulomb repulsion U and an effective charge-boson interaction[22], the HHH allows to reproduce the equilibrium optical properties of underdoped cuprates[23] and rationalize the presence of tail states responsible for the broad absorption below the CTP energy[27,31]. We argue that these tail states are responsible for the non-adiabatic contribution for sub-CTP pumping, i.e. for the long timescale "thermal" response displayed in Fig.2b. On the contrary the ultrafast transient optical response observed for $E_{pump}<E_{CTP}$ unveil a mixed regime of light-matter interaction where the light-driven gain of the electron kinetic energy leads to a sudden reaction of the bosonic field.

The Hubbard-Holstein Hamiltonian is given by the sum of three terms:

$$H_t = -t \sum_{i,\mu,\sigma} \left( e^{iA(\tau)} c^\dagger_{i+\mu,\sigma} c_{i,\sigma} + H.c. \right), \tag{1}$$

$$H_U = U \sum_i \left( n_{i,\uparrow} - \frac{1}{2} \right) \left( n_{i,\downarrow} - \frac{1}{2} \right), \tag{2}$$

$$H_{EPI} = \omega_0 \sum_i a^\dagger_i a_i + g\omega_0 \sum_i \left( a^\dagger_i + a_i \right)(1 - n_i), \tag{3}$$

accounting for the kinetic energy of the electrons (1), the onsite Coulomb repulsion (2), and the electron boson interaction (3), respectively. In equations 1-3 $t$ is the hopping amplitude, $c^\dagger_{i,\sigma}$ is the fermionic creation operator, $\mu$ is a unit vector along the axes of the lattice, $a^\dagger_i$ creates a boson at site $i$ with frequency $\omega_0$, and $n_i$ is the electron number operator. The units are such that $\hbar=1$. We choose model parameters typical for the cuprates ($t=0.29\,eV, \omega_0=0.2t, \lambda=\frac{g^2\omega_0}{4t}$). The value of the Hubbard repulsion $U=10t$ yields low-energy physics very similar to that of the $t-J$ model with $J=4t^2/U$ (see Methods). Although the exact solution for this Hamiltonian is not available, the time evolution of the many body wavefunction $|\Psi(\tau)\rangle$ can be obtained in a small two-dimensional lattice (8 sites with periodic boundary conditions) by using an exact diagonalization method (time dependent Lanczos approach) based on a smart truncation of the boson Hilbert space which has been proved successful in different systems[21,32,33]. Spin and charge degrees of freedom are exactly treated within numerical precision. The optical conductivity is then calculated at any value $\tau$ of the pump-probe delay, within the linear response theory, starting from the wave-function $|\Psi(\tau)\rangle$ (see Methods for more details).

In Fig.3b and Fig.3d we show the calculated $\Delta\sigma_1(\omega,\tau)$ at the delay time $\tau = 0.01$ ps when 20 fs long pump pulses are used for the calculations, and at larger time delay ($\tau = 0.04\,ps$). Small temporal length pulses have been employed for computational ease, and using longer pulses does not alter the emerging physical picture (see Methods). The pump energies are 3 eV (blue curves) and 1 eV (red curves), and the charge-boson coupling constant is $\lambda = 0.5$. By normalizing the probe delay over the pump pulse duration ($pd$), a fair comparison between experiments and theory can be performed (Fig.3a with Fig.3b at $\tau/pd = 0.5$, and Fig.3c with Fig.3d at $\tau/pd = 2$).

The major photo-induced feature for $E_{pump}$>$E_{CTP}$ (Fig.3b, Fig.3d) is a depletion of the optical conductivity which is almost independent on time. However we observe quantitative discrepancies between the linewidths calculated and the ones measured. In particular, the calculated drop of SW extends over a broader energy range. This is arguably due to the fact that the model is well suited to address doping phenomena[13] but lacks both the detailed crystal structure and a proper thermal reservoir. This is confirmed by comparing the expected variation of the optical conductivity upon doping[6] (black dashed curve inset Fig.1) with the model results (Fig. 3b). Hence, for long pump-probe delays or $E_{pump}$>$E_{CTP}$, the single boson mode contained in the HHH acts as an effective but quantitatively inaccurate thermal bath.

On the other hand, both theory and experiments for $E_{pump}$<$E_{CTP}$ display a spectral response exhibiting an evolution on ultrafast timescales, i.e. the depletion of SW at short pump-probe delays is centred at a slightly lower probe energy. We stress that such ultrafast changes of the spectral response are described by the model exclusively in presence of strong electron-boson interaction, as can be seen by inspection of the insets in Fig.3b ands Fig.3d that are obtained for $\lambda = 0$. The ultrafast "red-shift" of the optical response for $E_{pump}$<$E_{CTP}$ may be rationalized as the result of a polaronic distortion around holons and doublons leading to an energy renormalization whose estimate at $\lambda = 0.5$ is close to the experimental value 0.1 eV[34,35].

In the following we present the theoretical results in details. In order to confirm the assumption that inelastic (mixed) processes dominate in above-CTP (sub-CTP) cases we break up the wave function $|\Psi(t)\rangle = \sqrt{1 - |b(t)|^2}\,|\Psi_{GS}\rangle + b(t)|\phi(t)\rangle$ into the ground state $|\Psi_{GS}\rangle$ and the deformation component $|\phi(t)\rangle$, i.e. the part of the wave function different from the ground state. Since $|\phi(t)\rangle$ includes a doublon-holon pair at all characteristic times, $|b(t)|^2$ is a rough estimate of the double occupancy. Indeed, the number of doublons left in the system at $\tau$>0 is an order of magnitude smaller for sub-CTP pump energy with respect to the above-CTP one (Fig.4a), confirming that the inelastic processes are less relevant for sub-CTP pumping.

For highlighting the details of the changes of the ground state driven by the light pulses we singled out the perturbed component $|\phi(t)\rangle$ and calculated the expectation values of the different parts of the HHH. We calculated the electron kinetic energy, the boson number, and the interaction energy as the expectation values of eq.1 (apart the boson frequency), and of the first and second term of eq.3, respectively. While the first two terms have a self explanatory name, the interaction energy displayed in the inset of Fig.4a should be considered as a qualitative description of the state of the bosonic field. Naively, the more the bosonic field excitations are "localized" around the double occupancy the higher is the interaction energy.

For $E_{pump}$>$E_{CTP}$ a doublon-holon pair is created with a significant kinetic energy increase during the laser pulse and subsequently relaxes to lower energies emitting bosons on a longer time scale. We can see in Fig.4b that the decrease of the average kinetic energy is related to the increase of the average boson number (Fig.4b). The physics emerging after irradiation with high photon energy pulses is thus consistent with a thermalization scenario of the photo-excited doublons and holons.

For sub-CTP excitations a completely different response is observed. A heavily dressed quasi-particle is formed during the laser pulse and then relaxes on longer timescales, as shown in Fig.4c.

The average kinetic energy gain (negative on Fig. 4c) driven by the laser pulse is completely lost in a very short time. In the following femtoseconds, the kinetic energy variation turns positive (gray shaded area in Fig.4c), displays a maximum at $\tau_1^{model} = 25\,fs$, and finally relaxes at later times. This effect on the kinetic energy is accompanied by an increase of the absolute value of the charge-boson interaction energy, that exhibits its highest absolute value at $\tau_1^{model}$ marking a local coupling between bosons and electronic excitations (Fig.4a inset). These evidences indicate that the photo-excitation with energy smaller than the charge-transfer peak drives a collective response of the bosonic field resulting in the formation of massive quasi-particles which only at longer times evolve into a thermal incoherent state ($\tau_2^{model} > 40\,fs$). Such strong bosonic field distortion leads to the observed optical conductivity drop at the unperturbed charge-transfer edge which survives only in coincidence with the pump pulses. The time dependence of the average number of excited boson and the average kinetic energy confirm that the energy is transferred nearly "instantaneously" (on a timescales related to the inverse of the boson frequency) from the electromagnetic field into the boson subsystem for $E_{pump} < E_{CTP}$ only for non negligible electron-boson coupling (green line in Fig.4c).

The scenario described by means of ab-initio Hubbard Holstein calculations explains the different response measured for above-CTP and sub-CTP excitations (Fig.2). In particular, the strong bosonic field distortion leads to the quasi-instantaneous non-thermal quench of the optical response at the charge-transfer edge. It is interesting to note that for both theory and experiment the ultrafast response in the optical conductivity for sub-CTP excitation gives a transient feature which is qualitatively analogous to the changes associated to a small doping (Fig.1, black dashed curve inset). This observation leads us towards the speculation that a perturbation with sub-CTP photon energy drives a non-thermal tendency to delocalization that produces effects qualitatively analogous to the increased doping. This opens to several possible scenarios where coherent electromagnetic fields can be used to manipulate quantum coherent phases of matter[36,37].

**Methods**

**Experimental details.** We performed time-resolved experiments with broadband (single-colour) probe pulses and pump pulses with energies of 0.95±0.04 eV (1.4±0.2 eV) and 3.1±0.04 eV (2.3±0.3 eV) on $La_2CuO_4$. The durations of the pump pulses follows: $FHWM_{3.1eV} \sim 140$ fs, $FHWM_{0.95eV} \sim 100$ fs, $FHWM_{1.4eV} \sim 12$ fs, and $FHWM_{2.3eV} \sim 7$ fs. The broadband probe measurements source is a Ti:Sa amplified laser at 250 KHz of repetition rate, while the single-colour measurements have been performed with a Ti:Sa amplified source at 1 KHz. The sub-20 fs pulses are obtained by compressing with chirp mirrors (2.3±0.3 eV) or with two prisms (1.4±0.2 eV) the output of a tunable NOPA. The 0.95±0.04 eV pulses are generated with a commercial OPA, the 3.1±0.04 eV by second harmonic generation of the fundamental laser light in a 200 μm thick BBO crystal. As expected for self-phase modulation[38] the chirp of the broadband probes used in our experiments is mainly linear. Thus it has been simply corrected with the same straight line for both above-CTP and sub-CTP excitations. The pump diameters (w) are $w_{3.1eV} \sim 125$ μm, $w_{0.95eV} \sim 250$ μm, $w_{1.4eV} \sim w_{2.3eV} \sim 100$ μm. For the broadband probe measurements, freshly polished a-b oriented samples were mounted on the cold finger of a helium-flow cryostat. The reflectivity changes $\Delta R(\omega,\tau) = R(\omega,\tau)/R(\omega)-1$, with $R(\omega,\tau)$ time-dependent perturbed reflectivity and $R(\omega)$ equilibrium reflectivity, were measured in the 1.5-3.1 eV range at temperatures ranging from 330 K to 50 K. The pump fluences of the measurements shown in Fig.2a, Fig.2b, Fig.2c (not the inset), and Fig.3 are equal to 330 μJ/cm² and 170 μJ/cm² for the 3.1±0.04 eV and 0.95±0.04 eV pump pulses, respectively. These fluences correspond to about $4 \cdot 10^{-3}\,ph/u$ and $1 \cdot 10^{-3}\,ph/u$ for above-CTP and sub-CTP pump energies respectively, where $ph/u$ stands for photons ($ph$) absorbed by a chemical unit ($u$) of $La_2CuO_4$ i.e. by two lanthanum atoms, one copper, and four oxygens. The amplitude of the measured photo-induced variation of the reflectivity is linear with the pump intensity up to two times these values for both excitation wavelength (see Fig.S3 in Supplementary Information). The single-colour measurements shown in the inset of Fig.2c have been performed at room temperature with fluencies corresponding to ~2 mJ/cm$^2$ and ~5 mJ/cm$^2$ for above-CTP or sub-CTP pump pulses, respectively.

The long rise time observed for above-CTP pump might be partially due to the time taken by the photo-excited quasi-particles to reach the low energy states. It should be noted that the rise time detected with the ~140 fs long pump pulses at 3.1±0.04 eV (Fig.2a, Fig.2c without the inset) is longer than the one measured for resonant excitation of the CTP, i.e. with the ~7 fs long pump pulses at 2.3±0.3 eV (blue curve in Fig.2c inset). The rise time measured with $E_{pump}$=2.3±2.3 eV thus represents a lower boundary for the time a photo-excited charge takes to relax to the Hubbard gap and is consistent with the theoretical calculations (Fig.4b). We can exclude the slow rise time measured to be related to diffusion of the photo-excitation energy along the surface normal, as in such a case, we expect a different rise time for the different probe wavelength. In particular one would expect a longer dynamic for a smaller wavelength. This effect is not observed in our experiment, as can be seen in Fig.S4 in the Supplementary Information.

**Thermal effects.** The optical response of the charge-transfer edge as a function of pump-probe delay is reported in Fig.2c for visible (solid blue curve) and NIR pump pulses (solid red curve). The transient conductivity probed at 2.3 eV for 0.95±0.04 eV pump pulses is properly normalized by the factor $N$, calculated as follows. The pump energy density absorbed by the sample and detected by a single colour probe with energy $E_{probe}$ is equal to $\frac{\Phi(E_{Pump})}{x(E_{probe})} \cdot [1 - R(E_{Pump})] \cdot [1 - e^{-x(E_{probe})/x(E_{Pump})}]$, where $\Phi(E)$ is the fluence, $R(E)$ is the reflectivity, and $x(E)$ is the penetration depth for photon with energy $E$. Considering the bandwidth of the pump pulse and averaging over the probe energy range $\Delta E_{probe}$, the effective energy density $Q(E_{Pump}; E_{probe})$ is $Q(E_{Pump}; E_{probe}) = \frac{1}{\Delta E_{probe}} \iint \frac{\Phi(E_{Pump})}{x(E_{probe})} \cdot [1 - R(E_{Pump})] \cdot 1-e-x(Eprobe)/x(EPump)dEprobedEPump$. Making use of the reflectivity and the penetration depth measured via ellipsometry (see Fig.1 and Fig.S5b in the Supplementary Information), the ratio $N = \frac{Q(E_{Pump}=3.1\pm0.04eV; 1.9eV<E_{probe}<2.5eV)}{Q(E_{Pump}=0.95\pm0.04eV; 1.9eV<E_{probe}<2.5eV)}$ turns out to be ~14 at $T = 130\,K$, that is the normalization factor used in Fig.2c.

The absorbed pump energy density can be directly related to the increase of the sample temperature ΔT detected by the probe pulses. In particular $\Delta T = \frac{Q(E_{Pump}; E_{probe})}{\delta \cdot C}$ with $\delta$ density and $C$ the specific heat. Being $C_{mol} \approx 0.2\,J/K \cdot g$ at $T = 130\,K$[39] and $\delta \approx 7\,g/cm^3$, the expected temperature reached by the sample when all the degrees of freedom thermalize is $T + \Delta T \approx 139\,K$. Thus we expect that the laser-driven transient response at long pump-probe delays should be comparable to the difference between the equilibrium optical quantities measured at 130 K and at 139 K. This is indeed the case as can be seen by comparing the off-equilibrium Δσ$_1$(ω,τ=5 ps) with the difference between the equilibrium conductivities σ$_1$(T=139 K) and σ$_1$(T=130 K). See the black dashed curve in Fig.3c and Fig.S5 in the Supplementary Information.

**Data analysis.** In order to extract the time-dependent optical quantities such as the dielectric function (ε) or the optical conductivity (σ) out of pump-probed reflectivity ΔR(ω,τ)/R(ω)=R(ω,τ)/R(ω)-1, with R(ω) the equilibrium reflectivity and R(ω,τ) the pump-perturbed one, we proceed as follow. We obtained the equilibrium reflectivity R(ω) by measured ellipsometry data between 1 eV and 3.2 eV and from literature outside of this range[6,42]. We performed a Drude-Lorentz fit[40] of the broadband equilibrium reflectivity over a wide energy range, from few meV up to several tens of eV. We calculate the real and imaginary part of the equilibrium dielectric function, ε$_1$(ω) and ε$_2$(ω), on all energy range through the Kramers-Kronig (KK) relations[43].

The time-resolved ΔR(ω,τ)/R(ω) is measured in the 1.5-3.1 eV energy range. In order to obtain the time domain changes of the optical functions we consider small variations of the reflectivity ΔR(ω,τ) (up to 10$^{-2}$ in the measurement as shown in Fig.S1 of the Supplementary Information). We make the assumption that the variations of the pump-perturbed reflectivity outside of the probed spectroscopic region are either small (<10$^{-2}$) or too distant on the energy scale in order to significantly change the optical response in the probed optical range of 1.5-3.1 eV. Starting from this assumption both the measured reflectivity ΔR(ω,τ)/R(ω) and the equilibrium reflectivity R(ω) are known over a broad energy range, i.e. from few meV up to several tens of eV. Hence the time evolution of the optical quantities such as ε$_1$(ω,τ) and ε$_2$(ω,τ) can be obtained by the Kramers-Kronig transformations. The results of this analysis performed on the measurements at 130 K are reported in Fig.2 (besides Fig.2c inset). The photo-induced variation of the real part of the optical conductivity Δσ$_1$(ω,τ)=σ$_1$(ω,τ)-σ$_1$(ω), with σ$_1$(ω,τ) transient conductivity and σ$_1$(ω) the equilibrium one, is shown for pump energy equal to 3.1±0.04 eV (Fig.2a) and 0.95±0.04 eV (Fig.2b). The validity of this approach in our case is confirmed by comparing the obtained transient conductivities with the measured relative variation of the reflectivity (see Fig.2 in the main text and Fig.S1, Fig.S2 in the Supplementary Information).

We implement a transfer matrix method in order to correct the mismatch between the penetration depths of the pump and the probe pulses on the amplitude of the time-domain signal. We assume a pump-perturbed exponentially-graded index of refraction at the material surface by means of multiple 1 nm thick planes. The total reflection is calculated exploiting the continuity conditions for the electric fields across these boundaries. Further details can be found in ref.[44,45]. Being the penetration depth at ~0.95 eV larger than the one of any probe in the 1.5-3 eV range, one expects significative corrections only for above-CTP pump pulses (E$_{pump}$~3.1 eV). The detailed numerical analysis confirms this simple picture. This is made evident in Fig.S6,S7 of the Supplementary Information for pump-probe delay time τ equal to 0.1±0.1 ps. The ΔR(ω,τ)/R(ω) obtained in this way for E$_{pump}$~3.1 eV have been used as starting point for the Kramers-Kronig analisys described in the previous paragraph. The correction to transient reflectivity and dielectric functions are obtained, as summarized in Fig.S7 of the Supplementary Information.

**HHH details.** The time-dependent potential vector adopted is[46]

$$A = A_0 e^{-\frac{(\tau-\tau_0)^2}{\tau_{pump}^2}} \cos\left(\omega_{pump}(\tau-\tau_0)\right), \quad (4)$$

and the optical conductivity at time τ is given by[47]

$$\sigma(\omega,\tau) = \frac{1}{M\omega}\Im \int_0^\infty i e^{i(\omega+i\eta)t}\langle\Psi(\tau)|[j(t),j(0)]|\Psi(\tau)\rangle dt. \quad (5)$$

Here $|\Psi(\tau)\rangle = T e^{-i\int_0^\tau H(\tau_1)d\tau_1}|\Psi(\tau=0)\rangle$, $T$ is the time ordering operator, $j(t)$ is the current operator in the Heisenberg representation along one of the lattice direction axes, $\eta$ is a broadening factor taking into account additional dissipative processes, and $M$ is the number of lattice sites. The $|\Psi(\tau)\rangle$ state is obtained through the Lanczos time propagation method[48]. In the following we use periodic boundary conditions on 2D lattices[49] with $M=8$. The charge and spin degrees of freedom are evaluated exactly, whereas the quantum bosons are treated with the double boson cloud method[33]. The intensity of the classical vector potential $A(\tau)$ has been chosen such that the number of adsorbed photons per site during the pulse reproduces the experimental fluences.

In the paper we focused our attention on the properties of $|\phi(t)\rangle$, the photo-excited component of the wave-function $|\Psi(t)\rangle$. It is defined in terms of the wave-function at time $t$ and $|\Psi_{GS}\rangle$, the normalized ground state of the Hubbard-Holstein Hamiltonian in absence of the electromagnetic field: $|\Psi(t)\rangle = \sqrt{1-|b(t)|^2}|\Psi_{GS}\rangle + b(t)|\phi(t)\rangle$. Note that in Fig.4b and Fig.4c the kinetic energies and boson numbers shown are multiplied by the proper weighting factor $b(t)$. Moreover, in order to highlight small positive and negative variations of the kinetic energy (KE), we first multiply the modulus of KE by 125 and take the asinh(KE) function, which preserves the sign of the function and of its derivative (y-axis on the left in Fig.4b and Fig.4c).

The model calculations reported in the main text employed pump pulses with 20 fs duration. The variation of this quantity does not alter the physical picture as can be seen by comparing the ultrafast response for 20 fs, 40 fs, and 80 fs pumps in Fig.S8 of the Supplementary Information. A different spectral response with an optical conductivity drop that peaks at slightly lower probe energies for below-CTP pump pulses respect to above-CTP ones is present for all pulse durations. We stress that, as the pump pulses approach the experimental value of ~100 fs, there is a better agreement between the calculated spectral shift and the experimental one of ~0.1 eV.

[1] Imada, M., Fujimori, A., Tokura, Y. Metal-insulator transitions. Rev. Mod. Phys, **70**, 1039, 1998.

[2] Tokura, Y. Correlated-Electron Physics in Transition-Metal Oxides. Physics Today, **56**, 50, 2003.

[3] Baibich, M. N., et al. Giant Magnetoresistance of (001)Fe/(001)Cr Magnetic Superlattices. Phys. Rev. Lett. **61**, 2472, 1988.

[4] Grünberg, P. A. Nobel Lecture: From spin waves to giant magnetoresistance and beyond. Rev. Mod. Phys. **80**, 1531, 2008.

[5] Tokura, Y., Koshinara, S., Arima, T., Takagi, H., Ishibashi, S., and Uchida, S. Cu-O network dependence of optical charge-transfer gaps and spin-pair excitations in single-$CuO_2$-layer compounds. Phys. Rev. B **41**, 11657, 1990.

[6] Uchida, S. et al. Optical spectra of $La_2SrCuO_4$. Effect of carrier doping on the electronic structure of the $CuO_2$ plane. Phys. Rev. B **43**, 7942-7954, 1991.

[7] Perkins, J. D. et al. Mid-Infrared Optical Absorption in Undoped Lamellar Copper Oxides. Phys. Rev. B **71**, 1621-1624, 1993.

[8] Perkins, J. D. et al. Infrared optical excitations in $La_2NiO_4$. Phys. Rev. B **52**, R9863-R9866, 1995.

[9] Perkins, J. D. et al. Midinfrared optical excitations in undoped lamellar copper oxides. Phys. Rev. B **58**, 9390-9401, 1998.

[10] Mishchenko, A. S. et al. Charge Dynamics of Doped Holes in High Tc Superconductors: A Clue from Optical Conductivity. Phys. Rev. Lett. **100**, 166401, 2008.

[11] Lupi, S. et al. Far-Infrared Absorption and the Metal-to-Insulator Transition in Hole-Doped Cuprates. Phys.Rev. Lett. **102**, 206409, 2009.

[12] Nicoletti, D. et al. An extended infrared study of the p,T phase diagram of the p-doped Cu–O plane. New J. Phys. **13**, 123009, 2011.

[13] De Filippis, G. et al. Quantum Dynamics of the Hubbard-Holstein Model in Equilibrium and Nonequilibrium: Application to Pump-Probe Phenomena. Phys. Rev. Lett. **109**, 176402, 2012.

[14] Basov, D. N. et al. Electrodynamics of correlated electron materials. Reviews of Modern Physics, 83:471, 2011.

[15] Norman, M. R., Pepin, C. The electronic nature of high temperature cuprate superconductors. Rep. Prog. Phys. **66,** 1547, 2003.

[16] Han, J. E., Gunnarsson, O., and Crespi, V. H. Strong Superconductivity with Local Jahn-Teller Phonons in C60 Solids. Phys. Rev. Lett. **90**, 167006, 2003.

[17] Capone, M., Fabrizio, M., Castellani, C., Tosatti, E. Strongly Correlated Superconductivity. Science **296**, 2364, 2002.

[18] Bonča, J., Maekawa, S., Tohyama, T., Prelovšek, P. Spectral properties of a hole coupled to optical phonons in the generalized t-J model. Phys. Rev. B **77**, 054519, 2008.

[19] Cataudella, V., De Filippis, G., Mishchenko, A. S., Nagaosa, N. Temperature Dependence of the Angle Resolved Photoemission Spectra in the Undoped Cuprates: Self-Consistent Approach to the t-J Holstein Model. Phys. Rev. Lett. **99**, 226402, 2007.

[20] S. Dodge, S., Schumacher, A. B., Miller, L. L., Chemlaet D. S. Optically induced softening of the charge-transfer gap in Sr2CuO2Cl2. arxiv/0910.5048.

[21] Sangiovanni, G., Capone, M., Castellani, C., and Grilli, M. Electron-Phonon Interaction Close to a Mott Transition. Phys. Rev. Lett. **94**, 026401, 2005.

[22] In the paper we use the word boson in a broad sense, i.e. the bosonic mode can describe an electronic collective mode or one or more phonon modes. Nevertheless, "although the identity of this mode (be it an electronic collective mode or one or more phonon modes) remains controversial, the appearance of the dispersion renormalizations at multiple energy scales ranging from 10 - 110 meV strongly suggests coupling to a spectrum of oxygen phonons." arXiv:1306.2968v1.

[23] Okamoto, H. et al. Ultrafast charge dynamics in photoexcited Nd2CuO4 and La2CuO4 cuprate compounds investigated by femtosecond absorption spectroscopy. Phys. Rev. B **82**, 060513(R), 2010.

[24] Okamoto, H. et al. Photoinduced transition from Mott insulator to metal in the undoped cuprates Nd2CuO4 and La2CuO4. Phys. Rev. B **83**, 125102, 2011.

[25] Matsuda, K. et al. Femtosecond spectroscopic studies of the ultrafast relaxation process in the charge-transfer state of insulating cuprates. Phys. Rev. B **50**, 4097-4101, 1994.

[26] The spectral weigth is the integral of the optical conductivity. See e.g. Wooten, F., Optical properties of solids, Academic press, 1972.

[27] Falck, J. P., Kastner, M. A., Levy, A., and Birgenau, R. J. Charge-Transfer Spectrum and Its Temperature Dependence in La2CuO4. Phys. Rev. Lett. **69**, 1109-1112, 1992.

[28] Kim, Y. H. et al. Direct evidence of the importance of electron-phonon coupling in La2CuO4: Photoinducedir-active vibrational modes. Phys. Rev. B **36**, 7252-7255, 1987.

[29] Ginder, J. M. et al. Photoexcitations in La2CuO4: 2-eV energy gap and long-lived defect states. Phys. Rev. B **37**, 7506-7509, 1988.

[30] Masatoshi Imada, Atsushi Fujimori, and Yoshinori Tokura, Metal Insulator Transitions, Rev. Mod. Phys. 70, 1039 (1998).

[31] Lövenich, R., et al. Evidence of phonon-mediated coupling between charge transfer and ligand field excitons in Sr2CuO2Cl2. Phys. Rev. B **63**, 235104, 2001.

[32] Marchand, D. J. J., et al. Sharp Transition for Single Polarons in the One-Dimensional Su-Schrieffer-Heeger Model. Phys. Rev. Lett. **105**, 266605, 2010.

[33] De Filippis, G., Cataudella, V., Mishchenko, A. S., Nagaosa, N. Optical conductivity of polarons: Double phonon cloudconcept verified by diagrammatic Monte Carlo simulations. Phys. Rev. B **85**, 94302, 2012.

[34] Mishchenko, A. S., and Nagaosa, N. Electron-Phonon Coupling and a Polaron in the t-J Model: From the Weak to the Strong Coupling Regime. Phys. Rev. Lett. **93**, 36402, 2004;

[35] De Filippis, G., Cataudella, V., Mishchenko, A. S., and Nagaosa, N. Nonlocal Composite Spin-Lattice Polarons in High Temperature Superconductors. Phys. Rev. Lett. **99**, 146405, 2007.

[36] Fausti, D., et al. Light-Induced Superconductivity in a Stripe-Ordered Cuprate. Science **331**, 189, 2011.

[37] Kaiser, S., et al. Light-induced inhomogeneous superconductivity far above Tc in YBa2Cu3O6+x. arXiv:1205.4661

[38] Boyd, R. W. Nonlinear optics, Academic Press, third edition, 2007.

[39] Sun, K., et al. Heat capacity of single-crystal La2CuO4 and polycrystalline La1-xSrxCuO4 (0≤x≤0.20) from 110 to 600 K. Phys. Rev. B **43**, 239, 1991.

[40] Novelli, F., et al. Ultrafast optical spectroscopy of the lowest energy excitations in the Mott insulator compound YVO$_3$: Evidence for Hubbard-type excitons. Phys. Rev. B **86**, 165135, 2012

[41] Reul, J., et al. Temperature-dependent optical conductivity of layered LaSrFeO4. Phys. Rev. B **87**, 205142, 2013.

[42] McBride, J. R., Miller, L. R., and Weber, W. H. Ellipsometric study of the charge-transfer excitation in single-crystal La2CuO4. Phys. Rev. B **49**, 12224-12229, 1994.

[43] Wooten, F. Optical properties of solids, Academic press, 1972.

[44] Dal Conte, S., et al. Disentangling the Electronic and Phononic Glue in a High-Tc Superconductor. Science **30** 335 6076 1600-1603, 2012.

[45] Cilento, F. Non-equilibrium phase diagram of Bi2Sr2Y0.08Ca0.92Cu2O8+δ cuprate superconductors revealed by ultrafast optical spectroscopy. Ph.D. thesis, 2011.

[46] Matsueda, H., Sota, S., Tohyama, T., and Maekawa, S. Relaxation Dynamics of Photocarriers in One-Dimensional Mott Insulators Coupled to Phonons. J. Phys. Soc. Jpn. **81**, 013701, 2012.

[47] Fetter, A. L., and Walecka, J. D. Quantum Theory of Many Particle Systems, McGraW-Hill S. Francisco, 1971.

[48] Park, T. J., and Light, J. C. Unitary quantum time evolution by iterative Lanczos reduction. J. Chem. Phys. **85**, 5870, 1986.

[49] Dagotto, E., Joynt, R., Moreo, A., Bacci, S., Gagliano, E. Strongly correlated electronic systems with one hole: Dynamical properties. Phys. Rev. B **41**, 9049, 1990.

**Acknowledgements**

F.N. and D.F. are grateful to Marco Malvestuto for insightful discussions. The research leading to these results has been funded from the European Union Seventh Framework Programme [FP7/2007-2013] under project GO FAST grant agreement n° 280555. N.N. was supported by Grant-in-Aids for Scientific Research (Nos. 24224009) from the Ministry of Education, Culture, Sports, Science and Technology (MEXT) of Japan, Strategic International Cooperative Program (Joint Research Type) from Japan Science and Technology Agency, and by Funding Program for World-Leading Innovative R&D on Science and Technology (FIRST Program). M.C. and A.A. are financed by EU/FP7 through ERC Starting Grant SUPERBAD, Grant Agreement 240524.

**Author contribution statement**

F.N., D.F., and E.S. performed the broadband probe experiments. F.C. and D.F. developed the white-light optical setup. F.N., M.E., S.C., and G.C. performed the high-temporal resolution single-colour measurements. F.N. and D.F. performed the data analysis. M.G. and I.V.K. performed and analyzed magnetic and broadband ellipsometry measurements at several temperatures of interest. G.F., V.C., N.N., and A.M. developed the theoretical framework. G.F. and V.C. performed the calculations. F.N., D.F., G.F., and V.C. wrote the manuscript after discussing with all the co-authors. S.W., C.G., A.A., M.C., M.G. and F.P. have given particularly important comments. A.P. performed equilibrium reflectivity measurements in the visible range. D.P. and S.W. made and furnished high-quality oriented and polished single-crystals.

**Competing financial interests**

I declare that the authors have no competing interests as defined by Nature Publishing Group, or other interests that might be perceived to influence the results and/or discussion reported in this article.

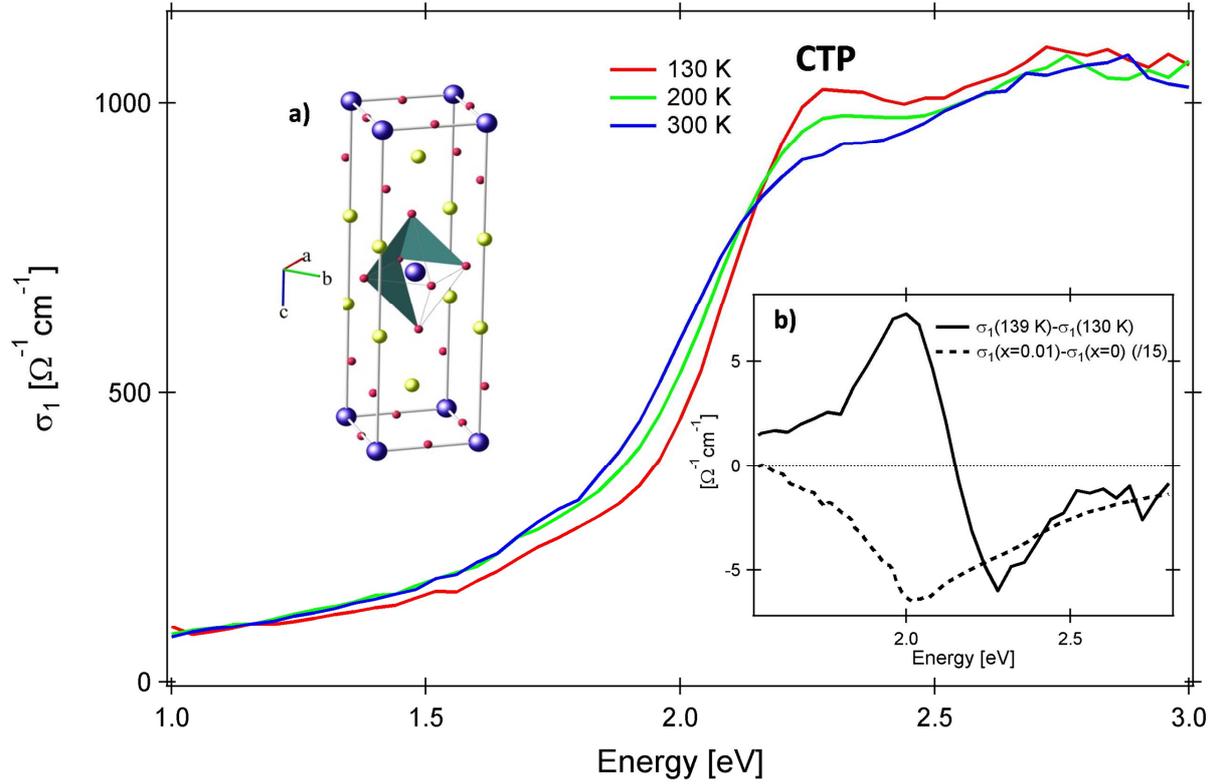

Figure 1. **Equilibrium optical conductivity** of lanthanum-copper oxide for polarization in the Cu-O (a-b) plane at 130 K, 200 K, and room temperature measured via ellipsometry. The charge-transfer optical peak (CTP), defined as the photon energy allowing for free-charge excitation, is indicated. a) $La_2CuO_4$ unit cell with copper in purple, oxygen in red, and lanthanum in yellow. b) Expected variation of $\sigma_1(\omega)$ upon doping (dashed line) or heating (solid line). The dashed line is the difference between the optical conductivity of $La_{1.99}Sr_{0.01}CuO_4$ (x=0.01) and $La_2CuO_4$ (x=0) at room temperature as obtained from ref.[6].

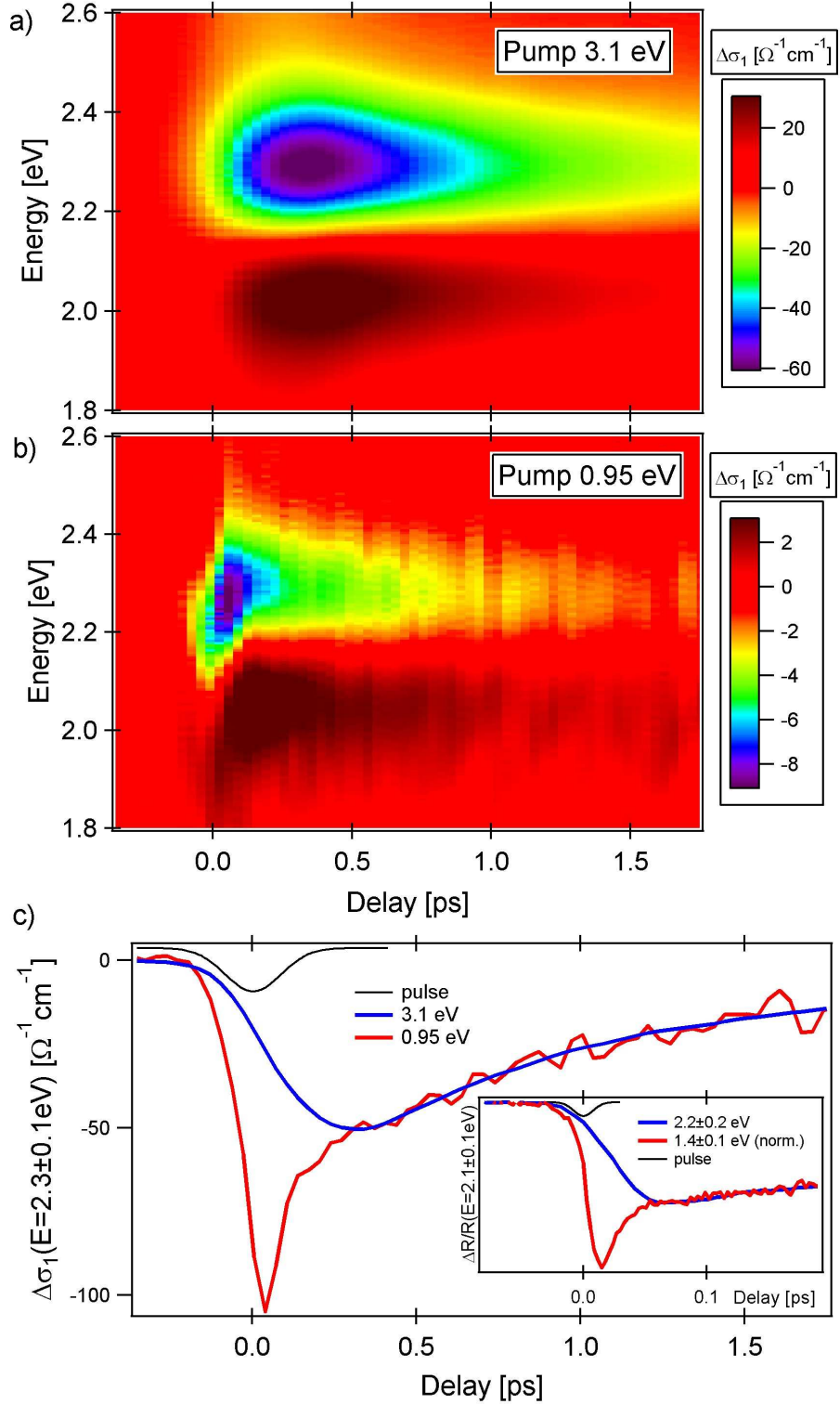

Figure 2. **Time-domain evolution of optical conductivity ($\Delta\sigma_1$).** The measurements performed at 130 K are reported as a function of probe energy for pump energy larger (3.1±0.04 eV, a) and smaller (0.95±0.04 eV, b) with respect to the charge-transfer peak. c) The transient optical conductivity at $E_{probe}$=2.3 eV for both pump energies is shown (3.1±0.04 eV in blue, 0.95±0.04 eV in red). The response for sub-gap excitation is multiplied by the ratio of the absorbed energy densities (see Methods). The black curve depicts the 3.1±0.04 eV pump autocorrelation. The inset of c) displays normalized single-colour pump-probe reflectivity measurements performed at room temperature with ~15 fs pulses (in black the pulse duration).

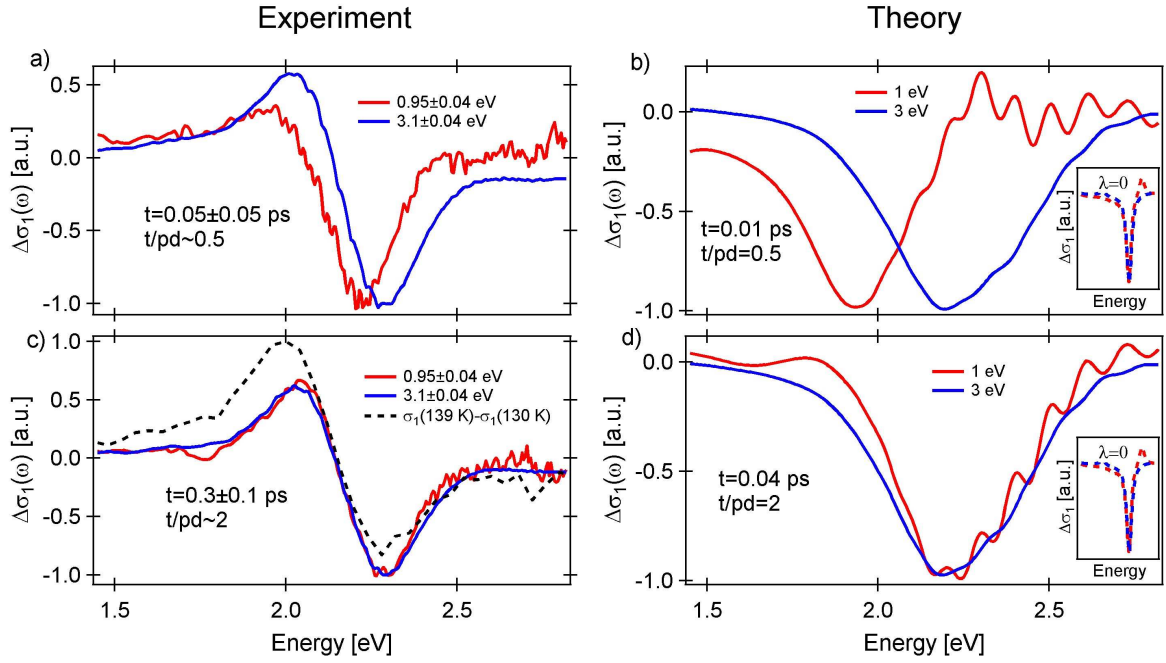

Figure 3. **Experimental and theoretical energy dependence of $\Delta\sigma_1$ at significative pump-probe delays**. In a) and c) the experimental data are shown, in b) and d) the model calculations. In the inset of b) and d) the zero-coupling results are reported. All the graphs share the same x axis. The data are normalized for the sake of comparison.

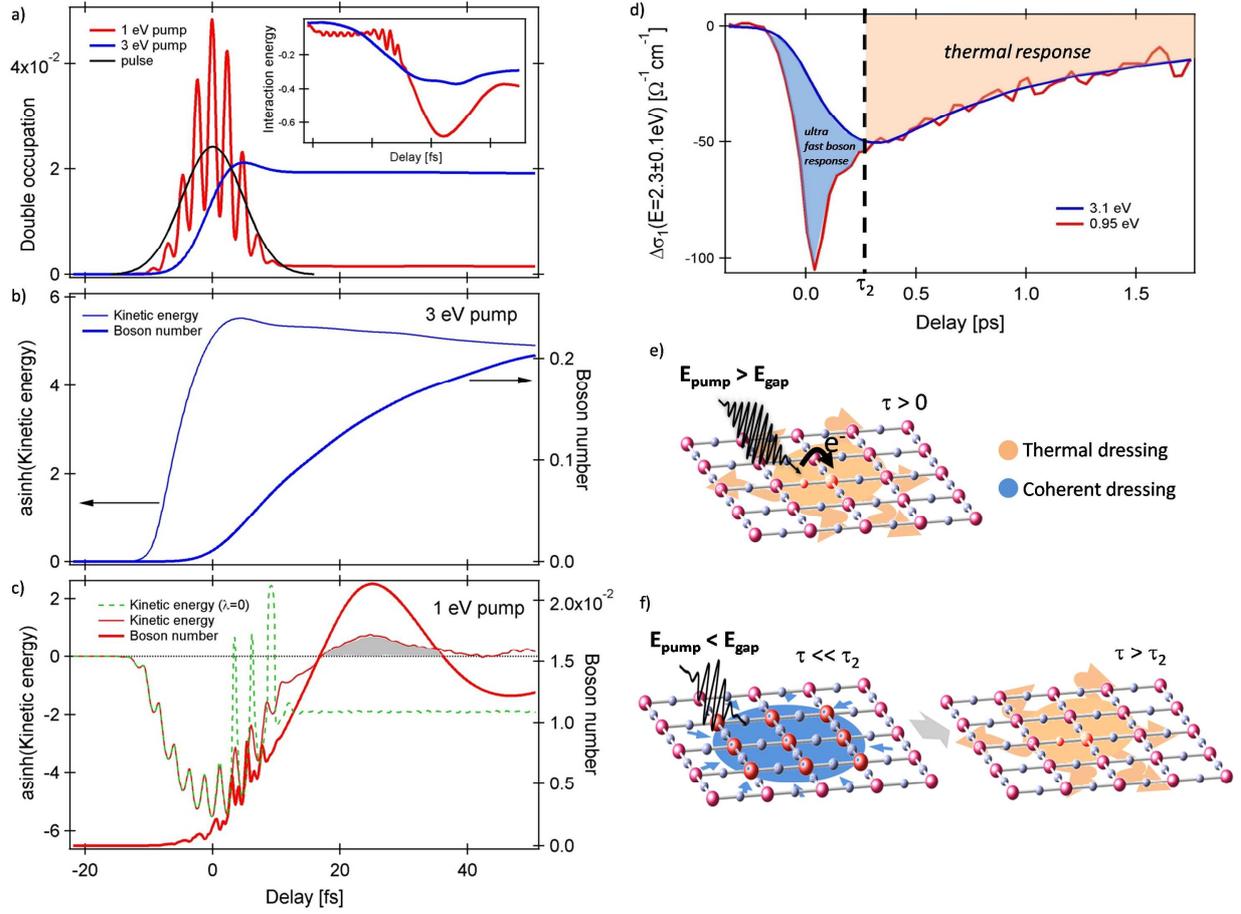

Figure 4. **Hubbard-Holstein calculations**. The average double occupancy (a) as a function of pump-probe delay for excitations above (blue) and below (red) the CTP. The average number of bosons (thick line) and the electron kinetic energy (thin line) for the two excitation wavelengths are reported in (b) and (c). In a), b), and c) the differences between the pump-perturbed quantities and the ground state ones are shown. In the inset of panel (a) the interaction energy is reported. Note that the kinetic energy axis in b) and c) is chosen to highlight both positive and negative small changes (see Methods). In d) Fig.2c is reproduced for clarity. Orange and blue shaded areas underline slow and fast responses of the bosonic field to resonant (e) and off-resonant (f) perturbations.